\documentclass[12pt,fleqn]{article}
\usepackage{amsfonts,amssymb}
\newcommand{\Bbc}[1]{{\mathbb #1}}
\newcommand{\End}{\nonumber\\}
\newcommand{\Real}{\Bbc{R}}
\newcommand{\Comm}{{}[\Brst,\chi]}
\newcommand{\Brst}{\Omega}
\newcommand{\BRST}{{\small BRST}}

\newcommand{\Gff}{\chi}
\newcommand{\Dpath}{{\cal{D}}}
\newcommand{\Intt}{\int_0^t}

\newcommand{\Intdt}{\int_0^{\delta t}}
\newcommand{\Pb}[2]{\{ #1,#2 \}}
\newcommand{\Dtwoh}[3]{\partial_{#2}\partial_{ #3}h(#1)}
\newcommand{\Dctwoh}[3]{D_{#2}D_{ #3}h(#1)}
\newcommand{\Dp}[2]{\frac{\partial #1}{\partial #2}}

\newcommand{\Xmu}{x^{\mu}}
\newcommand{\XXmu}{X^{\mu}}
\newcommand{\Pmu}{p_{\mu}}
\newcommand{\Vmu}{v_{\mu}}
\newcommand{\Vnu}{v_{\nu}}
\newcommand{\Tmu}{T_{\mu}}
\newcommand{\Etamu}{\eta^{\mu}}
\newcommand{\Pimu}{\pi_{\mu}}
\newcommand{\Lam}{\lambda}
\newcommand{\Kap}{\kappa}
\newcommand{\Gam}[3]{\Gamma_{#1 #2}{}^{#3}}
\newcommand{\Curv}[4]{R_{#1 #2 #3}{}^{#4}}
\newcommand{\Hamg}{H}
\newcommand{\Hamf}{H_f}
\newcommand{\Hamu}{H_u}
\newcommand{\Dadj}{\delta}
\newcommand{\Exp}[1]{\exp \left(  #1 \right)}
\newcommand{\Dh}[2]{\partial_{#2}h(#1)}

\newcommand{\Vip}{\,\iota\,}
\newcommand{\Fpos}{\theta}
\newcommand{\Fmom}{\rho}
\newcommand{\Fmpos}{\tilde{\Fpos}}
\newcommand{\Fmmom}{\tilde{\Fmom}}
\newcommand{\Bm}{\tilde{b}}

\newcommand{\xm}{\tilde{x}}

\newcommand{\Hstar}[1]{{}\,\,{}^* #1}

\newcommand{\Ch}{C_h}
\newcommand{\psit}{\tilde{\psi}}
\newcommand{\Meh}{M}
\newcommand{\Sigm}{\sigma_{\mu}}

\newcommand{\Frac}[2]{{\textstyle{\frac{#1}{#2}}}}
\newcommand{\Dut}{d_{u\,(2)}}
\newcommand{\Ga}{{{}_{[a]}}}
\newcommand{\Gb}{{{}_{[b]}}}
\newcommand{\Gab}{{}_{{}_{\Gamma_{ab}}}}
\newcommand{\Csh}{\frac{\cosh us}{\sinh  us}}
\begin{document}
\newtheorem{Def}{Definition}[section]
\newtheorem{The}{Theorem}[section]
\newtheorem{Lem}{Lemma}[section]
\bibliographystyle{unsrt}
\begin{flushright}
KCL-MTH-00-20
\end{flushright}
\begin{center}
{\LARGE The Topological Particle and Morse Theory}\\
 \ \\
  {\Large Alice Rogers}\\
   \ \\
    Department of Mathematics\\
    King's College\\
    Strand\\
    London WC2R 2LS \ \\
          May 2000
\end{center}
\begin{abstract}
Canonical \BRST\ quantization of the topological particle defined
by a Morse function $h$ is described. Stochastic calculus, using
Brownian paths which implement the WKB method in a new way
providing rigorous tunnelling results even in curved space, is
used to give an explicit and simple expression for the matrix
elements of the evolution operator for the \BRST\ Hamiltonian.
These matrix elements lead to a representation of the manifold
cohomology in terms of critical points of $h$ along lines
developed by Witten \cite{Witten82}.
\end{abstract}
%
%
\section{Introduction}
The topological particle, whose canonical \BRST\ quantization is
developed and applied in this paper, is the simplest example of a
topological quantum theory. There are many reasons, both physical
and mathematical, for studying such theories; however much of the
work to date has been carried out in the Lagrangian approach,
using the functional integral as the starting point. The
underlying motivation for this paper is the belief that a serious
canonical analysis of such theories should be fruitful.

The topological particle and its quantization is described by
Beaulieu and Singer in \cite{BeaSin}, but these authors
concentrate on the case based on a constant function on the
manifold, while in this paper the model is based on a Morse
function for the manifold, a function which is anything but
constant, having only isolated critical points, and encodes
information about the topology of the manifold. It is first shown,
in section \ref{CDsec}, that the the supersymmetric quantum
mechanical system which arises on \BRST\ quantization of the model
is the system used by Witten \cite{Witten82} in his work on
supersymmetry and Morse theory. The topological origin of this
model gives a natural explanation for the form of the matrix
elements for the theory which in Witten's paper are calculated by
instanton methods.

The main new result of the present paper is a path integral
formula for the calculation of these matrix elements, which is
derived in section \ref{PIsec} using the methods of stochastic
calculus on manifolds.  Paths are defined by a stochastic
differential equation which is essentially the Nicolai map for the
model \cite{Nicola80,BirRakTho88}; the paths encode fluctuations
about classical trajectories and thus lead to a fully rigorous
path integral WKB method (as derived by Blau, Keski-Vakkuri and
Niemi using physicists' methods \cite{BlaKesNie90}). The result
makes contact again with the classical action of the topological
particle which was the starting point.

Having established the precise form of the matrix elements of the
model, in section \ref{MTsec} these are used to construct
explicitly the cochains and cohomology for the model of the
manifold cohomology  introduced by Witten \cite{Witten82} based on
the critical points of  a Morse function.
\section{Classical Dynamics}
\label{CDsec}
The  topological particle model introduced by  Beaulieu and Singer
\cite{BeaSin} is defined by the action
 \begin{equation}\label{ACeq}
  S\left(x(.)\right) = \Intt \Vmu(x(t'))\, \dot{x}^{\mu}(t') \, d t'.
 \end{equation}
The fields $x$ are smooth maps  $x:I \to M$ from $I$
(the interval $[0,t]$ of the real line)
into  a compact $n$-dimensional Riemannian manifold $M$, while
 $v= d h$ is an exact one-form on $M$. The components $\Vmu$
in local coordinates $x^{\mu}, \mu = 1, \dots, n$ are as usual
defined by $v=\Vmu d \Xmu$, so that $\Vmu = \frac{\partial
h}{\partial \Xmu}$.

Clearly, since $v$ is the differential of $h$, this action can be
expressed more simply as
\begin{equation}\label{ACSeq}
  S\left(x(.)\right) = h(x(t)) - h(x(0)).
 \end{equation}
This form of the action shows that the model is indeed topological
in nature, a related point being that the equation of motion for
$x$ is trivially satisfied. However the form of the action
(\ref{ACeq}) involving positions and velocities is required for
the passage from the Lagrangian to the Hamiltonian form. While
Beaulieu and Singer considered the case $v=0$, in this paper a
more general situation is considered; in particular in section
\ref{MTsec} the function $h$ is taken to be a Morse function, that
is, a function on $M$ with isolated critical points.

It is evident that the action (\ref{ACeq}) is highly symmetric,
depending only on the endpoints of the path. It might thus be
naively supposed that the path integral
 \begin{equation}\label{NPIeq}
  \int_{\mbox{paths}/\mbox{symmetries}}
   \Dpath x(.) \exp \left( S(x(.))\right)
 \end{equation}
would be trivial. This is not in fact the case because the
`measure' $\Dpath x$ is not simply some limit of a product
measure, but must be derived by careful canonical quantization of
the theory, which is carried out below.

The first step in this process is to  investigate the classical
Hamiltonian dynamics. From the action (\ref{ACeq})
the Lagrangian of the theorem is seen to be
\begin{equation}\label{LAGeq}
  {\cal{L}}(x,\dot{x}) = \Vmu(x)\, \dot{x}^{\mu},
\end{equation}
so that the Euclidean time Legendre transformation to the phase
space $T^*M$ (the cotangent bundle of $M$) gives as momentum
conjugate to $\Xmu$
\begin{equation}\label{MOMeq}
  \Pmu = i\frac{\delta {\cal{L}}(x,\dot{x})}{\delta \dot{x}^{\mu}}
       = i\Vmu.
  \end{equation}
The symmetries of the system now manifest themselves as $n$ constraints
on the phase space $T^*M$:
 \begin{equation}\label{CONeq}
  \Tmu \equiv -\Pmu +i\Vmu(x) =0, \qquad \mu = 1, \dots, n.
 \end{equation}
The Poisson brackets on the phase space $T^*M$ are obtained from
the standard symplectic form $\omega=d\Pmu \wedge d\Xmu$, so that
as usual $\Pb{\Xmu}{p_{\nu}}= \delta^{\mu}_{\nu}$. Direct
calculation shows that (since $v$ is closed)
\begin{equation}
  \Pb{T_{\mu}}{T_{\nu}}=0.
\end{equation}
The Hamiltonian of the system is, by the Euclidean time
prescription,
 \begin{equation}\label{HAMCeq}
  H = i\Pmu \dot{x}^{\mu} + {\cal{L}}(x,\dot{x}) =0,
 \end{equation}
so that the constraints are first class and abelian.

As is standard in a topological theory, the constraints are of a
number that seems to preclude any interesting dynamics - in this
case the system has a $2n$-dimensional phase space with $n$ first
class constraints, so that by naive counting one would expect the
corresponding reduced phase space to be trivial. In fact  the
theory does capture some topological information as will emerge
below.

The first indication of this comes from considering gauge-fixing,
which shows that the reduced phase space, while as expected zero
dimensional, corresponds to the critical points of $h$. The
reduced phase space is defined to be the quotient of the subspace
of the phase space $T^*M$ on which the constraints hold by the
action of the group generated by the constraints
\cite{HenTei,KosSte}. Classically gauge-fixing conditions are
sought which pick out one point in each orbit of this group; in
this case a natural choice is the set of $n$ conditions $\XXmu
\equiv g^{\mu\nu}(-p_{\nu}-i\Vnu)=0$. (Justification for this
choice can only be fully made on quantization.) Taken together the
constraints and the gauge-fixing condition are satisfied when
$p_{\mu}=0$ and $\Vmu=0 $, that is, at the critical points of the
manifold. To see how this finite reduced phase space can provide
topological information we turn to quantization, using the \BRST\
approach.
\section{BRST quantization}
\label{BQsec}
To implement the constraints and gauge-fixing at the quantum level
we use the \BRST\ quantization in canonical form
\cite{Hennea,HenTei}, introducing ghosts and their conjugate
momenta. For this process two supermanifolds are required, a super
configuration space $SM$ with  even local coordinates $\Xmu$ and
odd local coordinates $\Etamu$ and a super phase space $SPM$ with
even local coordinates $\Xmu$ and $\Pmu$ and odd local coordinates
$\Etamu$ and $\Pimu$. (In each case the index ${\scriptstyle\mu}$
runs from $1$ to $n$.) The $(n,n)$ dimensional supermanifold $SM$
is built from the tangent bundle of $M$, with coordinate patches
corresponding to those on $M$ and changes of the coordinates
$\Xmu$ between patches being those on $M$ while those of the
coordinates $\Etamu$ are defined by
\begin{equation}\label{TRANSFUNCeq}
  \tilde{\Etamu}= \Dp{\tilde{x}^{\mu}}{{x}^{\nu}}\eta^{\nu}.
  \end{equation}
For future reference we note that there is a well-defined
projection $\epsilon:SM \to M$ defined by

\begin{equation}\label{Peq}
  \epsilon(x,\eta)=x.
\end{equation}

The super phase space $SPM$ is the cotangent bundle to $SM$, so
that $\Pmu$ and $\Pimu$ transform according to the rule
\begin{equation}\label{PTRANSFUNCeq}
  \tilde{\Pmu}= \Dp{x^{\nu}}{\tilde{x}^{\mu}}p_{\nu},\quad
  \tilde{\Pimu}= \Dp{x^{\nu}}{\tilde{x}^{\mu}}\pi_{\nu}.
\end{equation}

The simplest, and natural, choice of symplectic form on this
manifold, which makes $\pi_{\mu}$ the conjugate momentum to
$\eta_{\mu}$, is
 \begin{eqnarray}\label{SSYMPeq}
  w_s &=& d\left( \Pmu \wedge d\Xmu + \Pimu \wedge D\Etamu\right)\End
  &=& d \Pmu \wedge d\Xmu + D\Pimu \wedge D\Etamu
    -\frac12
    R_{\mu\nu\Lam}{}^{\Kap}\pi_{\Kap} \eta^{\Lam} d\Xmu \wedge dx^{\nu},
\end{eqnarray}
where the Levi-Cevita connection corresponding to the Riemannian
metric $g$ has been used, with Christoffel symbols
$\Gamma_{\mu\nu}{}^{\Kap}$ and curvature tensor components
$R_{\mu\nu\Lam}{}^{\rho}$, so that
\begin{equation}\label{CDIFeq}
  D\Etamu = d\Etamu + \Gam{\nu}{\Lam}{\mu} \eta^{\Lam} dx^{\nu},
 \quad D\Pimu = d\Pimu - \Gam{\nu}{\mu}{\Lam} \pi_{\Lam} dx^{\nu}.
\end{equation}
The corresponding Poisson brackets (which are calculated in
appendix \ref{PBap}) are:
\begin{eqnarray}\label{SPBeq}
  \Pb{p_{\nu}}{\Xmu} &=& \delta_{\nu}^{\mu},\quad \qquad
   \Pb{\Pmu}{p_{\nu}} =  R_{\mu\nu\Lam}{}^{\Kap}\pi_{\Kap}\eta^{\Lam}\End
  \Pb{\Pmu}{\eta^{\nu}} &=& \Gam{\mu}{\Lam}{\nu} \eta^{\Lam}, \quad
   \quad
  \Pb{\Pmu}{\pi_{\Lam}} = - \Gam{\mu}{\Lam}{\nu} \pi_{\nu}, \End
  \mbox{and} \quad \Pb{\pi_{\nu}}{\Etamu} &=& \delta_{\nu}^{\mu}.
\end{eqnarray}
the others being zero. To quantize, we take wave functions to be
functions $\psi(x,\eta)$ on the super configuration space $SM$.
The observables $\Xmu$ and $\Etamu$ are simply represented by
multiplication by these variables, while the momenta $\Pmu$ and
$\Pimu$ are represented as
\begin{equation}\label{SMOMeq}
  \Pmu = -i D_{\mu} \equiv -i \left( \Dp{}{x_\mu}
  + \eta^{\nu} \Gam{\mu}{\nu}{\Lam}\Dp{}{\eta^{\Lam}} \right)
  \quad \mbox{and} \quad
  \Pimu = -i\Dp{}{\Etamu}.
\end{equation}
The \BRST\ operator $\Brst$ is constructed from the constraints in
the standard way, giving
 \begin{equation}\label{BRSTDEFeq}
  \Brst =  \Etamu \Tmu =i\Etamu \left(\Dp{}{x^{\mu}} + \Vmu \right).
 \end{equation}
(The symmetry of the Christoffel symmetry removes the covariant
part of $p_{\mu}$ in this case, as in exterior differentiation of
forms.) The gauge-fixing fermion $\Gff$ is then constructed from
the gauge-fixing functions $\XXmu$ in the standard way:
\begin{equation}\label{GFFDEF}
  \Gff = \Pimu\Xmu =i g^{\mu\nu}\Pimu (D_{\nu}  - \Vnu).
\end{equation}
States of the system can of course naturally be identified with
forms on $M$ via the identification
\begin{equation}\label{FORMeq}
  a_{\mu_1 \dots \mu_p}(x) \eta^{\mu_1} \dots \eta^{\mu_p}
  \leftrightarrow
  a_{\mu_1 \dots \mu_p} (x)dx^{\mu_1} \dots dx^{\mu_p}
\end{equation}
Under this identification we see that
\begin{eqnarray}\label{NEWeq}
  \Brst &=&-i \Etamu\left(\Dp{}{x_\mu}-v_{\mu}\right)
  = i(d+ \Etamu\Vmu) =i e^h d e^{-h}, \End
   \Gff &=&-i g^{\mu\nu}\pi_{\nu}\left( \Dp{}{x_\mu}
  + \eta^{\nu} \Gam{\mu}{\nu}{\Lam}\Dp{}{\eta^{\Lam}} +\Vmu\right)=
   \Dadj - i g^{\mu\nu} \Vmu \Pimu =e^h \Dadj e^{-h}
\end{eqnarray}
where $d$ is exterior differentiation and $\Dadj$ is the adjoint
operator to $d$, that is $\Dadj={}^*d^*$ with ${}^*$ the Hodge
star operator. Thus we see that $\Brst$ and $\Gff$ are the
supersymmetry operators used by Witten in his study of
supersymmetry and Morse theory \cite{Witten82}. (The
identification of states with forms also leads to a natural inner
product on states; conventions for this may be found in appendix
\ref{SCap}.)

These expressions for $\Brst$ and $\Gff$  simplify the calculation
of the explicit expression for the canonical \BRST\ Hamiltonian
 $\Hamg  = -\frac{i}{2}\Comm$, leading to
\begin{equation}\label{BRSTHAMeq}
  \Hamg = \frac12 (d+\Dadj)^2
    +  \frac12 g^{\mu\nu} \Dp{h}{\Xmu} \Dp{h}{x^{\nu}}
    + \frac{i}{2}g^{\mu\Lam}(\eta^{\nu}\Pimu-\Pimu\eta^{\nu})
     \frac{D^2h}{Dx^{\Lam}Dx^{\nu}}
 \end{equation}
which is (up to a factor $\frac12$) the Hamiltonian used by Witten
\cite{Witten82}.  Witten also shows that the mapping $\psi \mapsto
e^{-h}\psi$ induces an isomorphism of de Rham cohomology classes
of $d$ and $\Brst =e^{-h}de^{h}$, and that forms with zero $H$
eigenvalue give exactly one representative of each $\Brst$
cohomology class. Since additionally $H$ has the same eigenvalues
as the Laplacian $d+ \Dadj$, we see that the gauge-fixing fermion
$\Gff$ is a good one \cite{GFBFVQ}.
\section{Path Integrals}
\label{PIsec}
In this section stochastic calculus is used to derive a rigorous
path integral expression for the action of the evolution operator
$\exp -\Hamg t$ on the states of the system.

In the special case of flat space (that is, where the manifold $M$
is simply $\Real^n$ with the Euclidean metric $g^{\mu\nu} =
\delta^{\mu\nu}$) this calculation was done by  Salmonson and van
Holten \cite{SalHol82} using WKB methods (also known as instanton
methods), which are a standard approximation technique in quantum
mechanics, \cite{LanLif58}. (An an accessible account of the
method applied to instantons may be found in the lectures of
Coleman \cite{Colema85}.) The basic idea is to consider
fluctuations about the classical trajectories. In the conventional
WKB approach only second order fluctuations are considered (first
order ones vanishing because the expansion is about the classical
trajectory) so that the method used is an approximate one; in this
paper we give an exact path integral formula in which the usual
WKB factor appears along with further factors. The approach is
valid on a manifold with a general Riemannian metric as well as in
flat space.

The stochastic calculus calculations which will now be given show
plainly how this arises. We will begin by working in flat space,
where the Hamiltonian is
 \begin{equation}\label{FLATHAMeq}
  \Hamf = \frac12 (d+\Dadj)^2
  +  \frac12  \Dp{h}{\Xmu} \Dp{h}{\Xmu}
    + \frac{i}{2}(\eta^{\nu}\Pimu-\Pimu\eta^{\nu})
     \frac{\partial^2h}{\partial\Xmu \partial x^{\nu}}.
 \end{equation}
One simple step will then adapt the method to a general Riemannian
manifold.

The  starting point is the stochastic differential equation
 \begin{equation}
  dx^{\mu}_t = db_t - \Vmu(x_t) dt, \qquad x_{0}=x
 \end{equation}
which implements the WKB approach of taking fluctuations about the
classical trajectories, and corresponds to the Nicolai map
\cite{BirRakTho88}. Here $x^{\mu}_t$ is a stochastic process on
the Wiener space of paths in $\Real^n$ starting from the point $x$
and $b^{\mu}_t$ is a standard Brownian path in $\Real^n$. Next,
for positive $t$, we consider the operator $U_t$ defined on
functions on the superspace $\Real^{n,n}$ by
 \begin{eqnarray} \label{UTeq}
  \lefteqn{U_t \psi{}(x,\eta) = } \End
&& \int  d\mu  \Big[  \Exp{\Intt  \left(\Dh{x_s}{\mu}dx^{\mu}_s
    +\Dtwoh{x_s}{\mu}{{\nu}}
  i   \Fpos^{\mu}_s \Fmom_{s\,\nu}ds\right)} \psi{}(x_t,\Fpos_t) \Big]\End
 \end{eqnarray}
 where $d\mu$ denotes Wiener measure for paths
 $(b^{\mu}_t,\Fpos_s^{\mu},\Fmom_{s \mu})$
 in superspace $\Real^{n,2n}$
 (for the fermionic paths $\Fpos_s$ and $\Fmom_s$ see \cite{JMPFI}).

Now by It\^o calculus,
 \begin{eqnarray}
  \lefteqn{ d\Big[\Exp{\Intt  \left(\Dh{x_s}{\mu}dx^{\mu}_s
    +\Dtwoh{x_s}{\mu}{{\nu}}
  i   \Fpos^{\mu}_s \Fmom_{s\,\nu}
  ds\right)}
    \psi{}(x_t,\Fpos_t) \Big]} \End
  =\Big[\exp&&\!\!
  \left(\Intt  \left(\Dh{x_s}{\mu}dx^{\mu}_s
    +\Dtwoh{x_s}{\mu}{{\nu}}
  i   \Fpos^{\mu}_s \Fmom_{s\,\nu}
  ds\right)\right)  (-\Hamf) \psi{}(x_t,\Fpos_t) \Big] dt \End
   &&+ \qquad \mbox{terms of zero measure},
 \end{eqnarray}
 so that
 \begin{equation}
  \frac{\partial U_t f(x)}{\partial t} = - U_t \Hamf f(x)
 \end{equation}
 and we conclude that
  \begin{equation}
 U_t = \exp -  t \Hamf.
 \end{equation}
 Now, again by It\^o calculus,
 \begin{equation}
 \Intt \left( \Dh{x_s}{\mu}dx^{\mu}_s
  + \frac12\Dtwoh{x_s}{\mu}{\mu} ds \right) =  h(x_t)- h(x)
 \end{equation}
 so that we can simplify (\ref{UTeq}) to obtain the Feynman-Kac-It\^o formula
  \begin{eqnarray}
 \lefteqn{\exp-t\Hamg \psi(x,\eta) = \int d\mu \Exp{-(h(x)-h(x_t))}} \End
 &&\Exp{
  \Intt \Dtwoh{x_s}{\mu}{{\nu}}
  i   \Fpos^{\mu}_s \Fmom_{s\,\nu}
     ds}  \psi(x_t,\Fpos_t).
   \end{eqnarray}
This expression shows  how the WKB factor $\Exp{-\Delta h}$ (which
clearly corresponds to the classical action (\ref{ACSeq}) of the
original topological theory) appears in the path integral.

The Feynman-Kac-It\^o formula is easily adapted to curved space
with a general Riemannian metric $g^{\mu\nu}$ by replacing the
Euclidean Brownian paths $b_t$ with the standard Brownian paths
$\Bm_t$ on a Riemannian manifold, and adjusting the fermion paths
by using the (stochastic) vielbein $e^{\mu}_{a,s}$ as specified
below. The bosonic Brownian paths on a Riemannian manifold, which
were introduced by Elworthy \cite{Elwort} and by Ikeda and Watanabe
\cite{IkeWat}, are defined by the stochastic differential equations
 \begin{eqnarray}
 d \Bm_t &=&  e^{\mu}_{a,t} db^{a}_t
 + \frac12 \Gam{\nu}{\rho}{\mu}(\Bm_t) dt   \End
  e^{\mu}_{a,t}&=&  \Gam{\nu}{\lambda}{\mu}(\Bm_t)  db^{b}_t
   + \frac12 e^{\nu}_{a,t} \Curv{}{}{\nu}{\mu}(\Bm_t) dt \End
 &&  \Bm_0^{\mu} = x^{\mu}, \quad e^{\mu}_{a,0} = e^{\mu}_{a}(x)
   \end{eqnarray}
where $x$ is the point on the manifold from which the Brownian
motion is chosen to start and $ \{e_a = e^{\mu}_{a}(x)
\Dp{}{\Xmu}, a=1 \dots n\}$ is a choice of orthonormal basis at
that point. The fermionic paths $\Fmpos_t^{\mu},\Fmmom_{t,\nu}$
are obtained from the flat space fermionic paths by rotating with
the stochastic vielbein:
 \begin{eqnarray}
   \Fmpos_t^{\mu} &=&  \Fpos^{a}_t e^{\mu}_{a,t} \End
   \Fmmom_{t,\nu} &=&  \Fmmom_{a\,t}e^{\mu}_{a,t} g_{\nu\mu}(\Bm_t).
 \end{eqnarray}
Using paths $\xm_t$ on $M$ satisfying
 \begin{eqnarray}\label{SDMeq}
  d\xm^{\mu}_t &=& d\Bm_t -g^{\nu\mu}(\xm_t) v_{\mu}(\xm_t) dt     \End
  \xm_0  &=& x,
 \end{eqnarray}
similar steps to those above lead to the Feynman-Kac-It\^o formula
   \begin{eqnarray}\label{FKIeq}
 \lefteqn{\exp-t\Hamg \psi(x,\eta) = \int d\mu \exp( - (h(x)-h(\xm_t))) } \End
 &&
 \exp \Big(\Intt \big( \Dctwoh{\xm_s}{\mu}{\nu}
  i   g^{\lambda\nu}(\xm_s)\Fmpos^{\mu}_s \Fmmom_{s\,\lambda}
   \End
 &+& \Curv{}{}{\mu}{\nu}(\xm_s) \Fmpos^{\mu}_s \Fmmom_{\nu\, s}
 +\Frac12 R_{\mu\kappa}^{\lambda\nu}(\xm_s)
 \Fmpos^{\mu}_s \Fmpos^{\kappa}_s \Fmmom_{\lambda\, s}\Fmmom_{\nu\, s}
          \big)ds \Big) \psi(\xm_t,\Fmpos_t).
   \end{eqnarray}
In both cases care must be taken when $x$ is a critical point,
since there will not in general be a unique solution to the
stochastic differential equation concerned.
\section{Morse theory and cohomology}
\label{MTsec}
In \cite{Witten82} Witten rescales  the function $h$  by a
constant factor (here called $u$) to obtain the scaled Hamiltonian
  \begin{equation}\label{SHAMeq}
  \Hamu = \frac12 (d+\Dadj)^2 + u^2 g^{\mu\nu} \Dp{h}{\Xmu} \Dp{h}{x^{\nu}} +
 u \frac{i}{2}g^{\mu\Lam}(\eta^{\nu}\Pimu-\Pimu\eta^{\nu})
     \frac{D^2h}{Dx^{\Lam}Dx^{\nu}},
 \end{equation}
and, taking the large $u$ limit, builds an explicit model for the
cohomology of the manifold in terms of the critical points of the
manifold with differential derived from the exterior derivative.
This leads directly to the weak and strong Morse identities for
$M$, but also gives considerably further insight into the mechanism
relating the critical points to the manifold topology. Some parts
of Witten's analysis have been proved rigorously, (for instance by
Bismut \cite{Bismut86} and by Simon et al \cite{Simetal}); however
the explicit modelling of the manifold's cohomology via critical
points and instanton calculations does not appear to have received
a full mathematical treatment of the nature given below.

For the rest of this paper it will be assumed that $h$ is a Morse
function, that is, it has only isolated critical points. (For
terminology and notation see appendix \ref{MTap}.) If $a$ is a
critical point of $h$ with index $p$  then Witten \cite{Witten82}
shows that for large $u$ there is exactly one  $p$-form
$\psi_{(u),a}\,(x,\eta)$ on $M$  which is concentrated near $a$
and  is an eigenstate of $\Hamu$ with eigenvalue $\lambda_a(u)$
which is low, that is, which does not grow like $u$ but is $
{{o}}(u)$. (This result is derived analytically by Simon et al in
\cite{Simetal}.) Additionally it is shown that there are no other
forms which have low $\Hamu$ eigenvalues. Witten also shows (as
was remarked in section \ref{BQsec} for the $u=1$ case) that the
mapping $\psi \mapsto e^{-hu}\psi$ induces an isomorphism of de
Rham cohomology classes of $d$ and $d_u=e^{-uh}de^{uh}$, and that
forms with zero $\Hamu$ eigenvalue give exactly one representative
of each $d_u$ cohomology class.

Observing that $d_u$ and $\Hamu$ commute, we see that if $\psi$ is
an eigenstate of $\Hamu$, then $d_u \psi$ is either zero or an
eigenstate of $\Hamu$ with the same eigenvalue. Thus if $a$ is a
critical point of $h$ with index $p$,
 \begin{equation}\label{CBeq}
 d_u \psi(a) = \sum_{b \in \Ch, {\rm index\ of\ }b=p+1}
 c_{ab} \psi_{b}
 \end{equation}
for some real numbers $c_{ab}$. As a result the cohomology of $M$
can be modelled by $p$-cochains
 (with $p=1,\dots,n={\rm dim}\, M$) of the form
 \begin{equation}
   c= \sum_{a \in \Ch, {\rm index\ of\ }a=p} c_a\psi_{(u),a}
 \end{equation}
where the coefficients $c_a$ are real numbers, with the modified
exterior derivative $d_{u}$ as coboundary operator. The calculation
of $c_{ab}$ in Witten's paper is done by instanton methods, which
may be made both more rigorous and more transparent by using the
path integral expression for $\exp -\Hamu t$ developed in the
preceding section, as will now be described.

The constants $c_{ab}$ in equation(\ref{CBeq}) may be evaluated by
considering the matrix elements $\Dut\exp -\Hamu t(A,B)$ in the
large $u$ limit. (Here the notation $\Dut$ means that the operator
$d_u$ acts with respect to the second argument.) To see that these
matrix elements are relevant, we choose an orthonormal basis of
eigenstates of $\Hamu$ consisting of the low eigenvalue states
$\psi_c, c \in \Ch$ (where we have simplified the notation by
dropping explicit reference to $u$) together with further
eigenstates $\{\psi_n|n=0, \dots, \infty\}$ with eigenvalues
$\Lam_{n}(u)$ which will be at least of order $u$. We can then
express the matrix elements of the evolution operator as
\begin{equation}
  \exp -\Hamu t(Y,X) =\sum_{c \in \Ch}^{\infty}
  e^{-\Lam_c(u) t}\Hstar{\psi_c(Y)}\, \psi_c(X) \quad + \quad
  \sum_{n=0}^{\infty} e^{-\Lam_n t}\Hstar{\psi_n(Y)}\, \psi_n(X).
\end{equation}
For large $u$ we have an approximate expression
 \begin{equation}
  \exp -\Hamu t (Y,X) = \sum_{c \in \Ch}^{\infty}
  \Hstar{\psi_c(Y)}\, \psi_c(X),
 \end{equation}
so that we have at leading order for large $u$
 \begin{equation}
  \Dut \exp -\Hamu t (Y,X) = \sum_{c \in \Ch}^{\infty}
  \Hstar{\psi_c(Y)}\, d\psi_c(X).
 \end{equation}
Now as was remarked above, each $\psi_c$ is concentrated around
$c$.  We thus expect that at leading order for large $u$
 \begin{equation}\label{CVALeq}
  \Dut \exp -\Hamu t (A,B) =   c_{ab} \Hstar{\psi_a(A)}\, \psi_b(B),
 \end{equation}
if $\epsilon(A)=a$ and $\epsilon(B)=b$.

In order to evaluate this expression we make use of the
Feynman-Kac-It\^o formula (\ref{FKIeq}), together with the explicit
form of the kernel in the neighbourhood of a critical point. We
choose a metric which globally satisfies the Smayle transversality
condition for $h$ (see appendix \ref{MTap}), and additionally one
which is Euclidean within the Morse coordinate neighbourhood $N_a$
of each critical point $a$ and on a neighbourhood of each steepest
descent curve $\Gamma_{ab}$ joining critical points.

Before proceeding further it is useful to introduce some specific
coordinate systems. For each critical point $c$ in $M$ we will
choose on $N_c$ a fiducial set of Morse coordinates (appendix
\ref{MTap}) $x_{[c]}^{\mu}$ and fermionic partners
$\eta_{[c]}^{\mu}$. Additionally for each steepest descent curve
 $\Gamma_{ab}$ (satisfying (\ref{STeq})) joining the pair of
critical points $a$ and $b$ with indices $p$ and $p+1$ respectively
we will choose a coordinate neighbourhood $U\Gab$ which contains
 $N_a\cup N_b \cup U\Gab$  with coordinates
$x\Gab,\eta\Gab$  such that $\Gamma_{ab}$ lies along $x\Gab^{n-p}$
while $x\Gab,\eta\Gab$ match $x\Gb,\eta\Gb$ on $N_b$ apart from
possible rotations, and also match  $x\Ga$ on $N_a$ apart from
possible rotations and  (necessarily) a translation in the
$x^{n-p}$ coordinate with $x^{n-p}\Gab=x\Ga^{n-p} + k_a$ for some
positive constant $k_a$. Reconciliation with the fiducial
coordinates will ultimately bring in sign factors.

Within $N_a$ the Hamiltonian then has the form
\begin{eqnarray}
  \Hamu &=& \sum_{\mu=1}^n \Bigg[
   \Frac12 \Big(-\Frac{\partial^2}
   {\partial x\Gab^{\mu} \partial x\Gab^{\mu}}    +
   u^2 (x\Gab^{\mu}-a\Gab^{\mu})(x\Gab^{\mu}-a\Gab^{\mu})\Big)
    \End
 &&\qquad \quad + \quad \Frac{i}{2}u \Sigm
    (\eta\Gab^{\mu}\pi\Gab{}_{\mu}-\pi\Gab{}_{\mu}\eta\Gab^{\mu}) \Bigg],
\end{eqnarray}
where
 $\Sigm=1, \mu=1, \dots n-p$ while
 $\Sigm=-1, \mu=n-p+1, \dots n$.
 The bosonic and fermionic parts commute so that their heat
kernels may be considered separately; the bosonic part is the
Harmonic oscillator Hamiltonian whose heat kernel is given by
Mehler's formula \cite{Simon}, while the fermionic part is (apart
from sign factors $\Sigm$) the fermionic oscillator whose heat
kernel is given in \cite{GBM}. If $x$ is near $a$ and in $N_a$ then
at leading order for large $u$
 \begin{eqnarray}
  \lefteqn{ \exp -\Hamu{}t (A,X) =_{{\rm def}} \Meh(A,X) } \End
  &=&      \left( \frac{u}{\pi}\right)^{n/2}
     \Exp{- \Frac12 u (x\Gab{}-k_a)^2}
     \prod_{\mu=1}^{n-p}( - \alpha\Gab^{\mu})\prod_{\nu=n-p+1}^{n}
      \eta\Gab^{\nu}  \End
 \end{eqnarray}
where $X$ is a point over $x$ in $N_{a}$, with  coordinates
$(x\Gab,\eta\Gab)$.

Next we calculate $\exp -\Hamu t(A,X)$ for $x$ near $\Gamma_{ab}$
using the Feynman-Kac-It\^o formula (\ref{FKIeq}). In this case
the steepest descent  curve (satisfying (\ref{STeq})) from $x$
approaches $a$ very fast. Thus after very small time $\delta t$
the path $\tilde{x}_{\delta t}$ is almost certainly near $a$, so
that to leading order in $u$ we have a contribution from
$\Gamma_{ab}$ of
  \begin{eqnarray}\label{GABKEReq}
  \lefteqn{ \exp -\Hamu t(A,X)\Gab{}=\exp -\Hamu \delta t
  \exp -\Hamu (t-\delta t) (A,X) }  \End
  &=&  \int d\mu  \Bigg[ \Exp{- u( h(x)-h(\xm_{\delta t}))} \End
\times && \exp\Bigg( \Intdt  u\Dctwoh{\xm_s}{\mu}{{\nu}}
  i g^{\lambda\nu}(\xm_s)  \Fmpos^{\mu}_s \Fmmom_{s\,\lambda}
   \End
&& + \Curv{}{}{\mu}{\nu}(\xm_s) \Fmpos^{\mu}_s \Fmmom_{\nu\, s}
 +\frac12 R_{\mu\kappa}^{\lambda\nu}(\xm_s)
 \Fmpos^{\mu}_s \Fmpos^{\kappa}_s \Fmmom_{\lambda\, s}\Fmmom_{\nu\, s}
          \,ds \Bigg)
 \Meh(a\Gab,\alpha\Gab,\xm_{\delta t},\Fmpos_{\delta t}) \Bigg]  \End
  &=& \left( \frac{u}{\pi}\right)^{n/2}\Exp{ - u(h(x)-h(a))}
   \prod_{\mu=1}^{n-p}( - \alpha\Gab^{\mu})\prod_{\nu=n-p+1}^{n}
      \eta\Gab^{\nu}. \End
 \end{eqnarray}
Here we have used the fact that the operator $\Etamu\Pimu$ which
corresponds to the term $g^{\lambda\nu}(\xm_s)  \Fmpos^{\mu}_s
\Fmmom_{s\,\lambda} $ in the path integral has zero eigenvalue on
 $\exp-\Hamu t (A,\xm_{\delta t},\Fmpos_{\delta t})$ when $x$ lies on $\Gamma_{ab}$.

To calculate $\Dut \exp -\Hamu t(A,B)$ we cannot take the
derivative of the separate contributions from each $\Gamma_{ab}$
using (\ref{GABKEReq}) because as we vary $x$ around $b$ we will
jump from one $\Gamma_{ab}$ to another. To avoid this difficulty we
note that
  \begin{eqnarray}
  \lefteqn{\Dut \exp -\Hamu t (A,B) }\End
  &=& \Dut\exp-\Hamu s \,  \exp -\Hamu (t-s)(A,B)  \End
   &=& \int_{M} d^n x d^n \eta   \exp-\Hamu (t-s) (A,X) \Dut \exp -\Hamu s(X,B).
  \label{FACeq}\end{eqnarray}
Because of the concentration of $\Dut\exp -Hs(X,B)$ near $b$ we
can integrate over $\Real^n$ rather than $M$ using the form of
$\exp -Hs(X,B)$ which is approximately true for large $u$ on
$N_b$; although ultimately we will obtain a result independent of
$s$ and $t$, at this stage we must use Mehler's formula in full
(including terms of order $e^{-us}$ whose equivalent we could
neglect near $a$ for our purposes) because it is not the zero mode
of $\Hamu$ which will contribute to $d\psi_a(b)$ at leading order.
Thus for $x$ and $y$ near $b$ we use
 \begin{eqnarray}
 \lefteqn{\exp -\Hamu s(X,Y) = \left( \frac{u}{\pi}\right)^{n/2}
 \Exp{-  \Frac12 u \left(x\Gab^2 \Csh \right) +u \,\frac{x\Gab y\Gab}{\sinh us}} }\End
 &&\qquad\qquad\times \,\prod_{\mu=1}^{n-p-1}
  \left( \phi\Gab^{\mu} e^{- us} - \eta\Gab^{\mu}  \right)
 \prod_{\nu=n-p}^{n} \left( \phi\Gab^{\nu}  - \eta\Gab^{\nu}e^{- us}
 \right).
 \End
 \end{eqnarray}
where and $(x\Gab,\eta\Gab)$, $(y\Gab,\phi\Gab)$ are the
coordinates of $X$ and $Y$ respectively, so that at leading order
in $u$ the relevant term of $\Dut\exp -\Hamu s(X,B) $ (that is, the
term which contains $d\psi_a(b)$)is
 \begin{equation}
 \left( \frac{u}{\pi} \right)^{n/2}
 u\,\frac{x^{n-p}}{\sinh us} \Exp{-  \Frac12 u \left(x\Gab^2 \Csh \right)}
  \quad\,\prod_{\mu=1}^{n-p} \eta\Gab^{\mu}
 e^{-us} \prod_{\nu=n-p}^{n}  \beta\Gab^{\nu}  .
  \end{equation}
Using (\ref{FACeq}) we see that
 \begin{eqnarray}
\lefteqn{\Dut  \exp -\Hamu t (A,B)}\End
 &=&\int_{\Real^n} d^{n}x\Gab \,
 \left(\frac{u}{\pi}\right)^{n}\theta(x\Gab^{n-p})
 u\, \frac{e^{-us}}{\sinh us} x\Gab^{n-p}
 \prod_{\mu=1}^{n-p} \left(  - \alpha\Gab^{\mu} \right)
 \prod_{\nu=n-p}^{n}  \beta\Gab^{\nu}\End
 &&\times\Exp{-\Frac12 u x\Gab^2 \Csh}\exp-u(h(x)-h(a)) \End
 &=&
 \int_0^{\infty} dx\Gab^{n-p}
 \left(\frac{u}{\pi}\right)^{(n+1)/2}u\, \frac{e^{-us}}{\sinh us} x\Gab^{n-p}
 \prod_{\mu=1}^{n-p} \left(  - \alpha\Gab^{\mu} \right)
 \prod_{\nu=n-p}^{n}  \beta\Gab^{\nu}\End
  &&\times \Exp{-\Frac12 u x\Gab^2 \left(\Csh-1 \right)}\exp-u(h(b)-h(a)) \End
 &=&\left(\frac{u}{\pi}\right)^{(n+1)/2}
 \prod_{\mu=1}^{n-p} \left(  - \alpha\Gab^{\mu} \right)
 \prod_{\nu=n-p}^{n}  \beta\Gab^{\nu}
 \exp-u(h(b)-h(a)) \End
 && \mbox{\ at leading order in $u$}.
 \end{eqnarray}
Here the $\theta$-function occurs because the contribution from
$\Gamma_{ab}$ to $\exp -\Hamu (t-s)$ is zero on the side of $b$
away from $a$. Now using equation (\ref{CVALeq}) and the fact that
 \begin{equation}
 \Hstar\psi_{a}(A)= \left( \frac{u}{\pi}\right)^{n/4}
 \prod_{\mu=1}^{n-p}\alpha\Ga^{\nu},
  \quad
 \psi_{b}(B)= \left( \frac{u}{\pi}\right)^{n/4}
 \prod_{\nu=n-p}^{n}\beta_{[b]}^{\nu}
  \end{equation}
we see that
 \begin{equation}
 c_{ab} = \left( \frac{u}{\pi}\right)^{1/2}
   \exp-u(h(b)-h(a))\sum_{\Gamma_{ab}} (-1)^{\sigma\Gab}
 \end{equation}
where $(-1)^{\sigma\Gab}$ is a sign factor which comes from the
change between the $[a]$ and $[b]$ coordinates and the
$[\Gamma_{ab}]$ coordinates.

If (once again following Witten \cite{Witten82}) we rescale each
$\psi_c$ to $\psit_c=e^{-uh(c)}\psi$, and additionally use
$\tilde{d_{u}}= \sqrt{\frac{\pi}{u}}d_u$, we obtain
 \begin{equation}
  \tilde{d}_u\psit_a
  = \sum_{\Gamma_{ab}} (-1)^{\sigma_{\Gamma_{ab}}} \psit_{b}
  \end{equation}
which coincides with the geometrical approach using ascending and
descending \break spheres.
\section{Conclusions and further possibilities}
 \label{CFsec}
In this paper we have carried out  a full canonical quantization
of the simplest topological quantum theory, the topological
particle, and demonstrated precisely the way in which the
quantization captures topological information. The path integral
formula developed in Section \ref{PIsec}, which implements the WKB
approximation in a mathematically rigorous way, even in curved
space, should be useful in other situations involving quantum
tunnelling.

Recent work by Hrabak \cite{Hrabak} on the BRST operator for the
two-dimensional topological sigma model leads (in an elegant and
original way using the multi-symplectic formalism) to the
supersymmetric theory considered and exploited by Witten
\cite{Witten88}. A novel approach to quantization in the
multi-symplectic formalism has been developed by Kanatchikov
\cite{Kanatc} which might make possible a new approach to
quantization of the two dimensional model.
 \vskip 1cm \noindent {\Large \bf Appendix}
\appendix
\section{Poisson brackets}
\label{PBap}

To calculate the Poisson brackets of the canonical variables
$\Xmu$, $\Pmu$, $\Etamu$ and $\Pimu$ determined by the symplectic
form (\ref{SSYMPeq}) on the super phase space $SPM$ introduced in
section \ref{BQsec} we first need the Hamiltonian vector fields of
these variables. The Hamiltonian vector field $X_f$ of a function
$f$ on phase space is defined by
\begin{equation}\label{HVFeq}
  X_f \Vip \omega_s = d f,
\end{equation}
where $\Vip$ denotes interior product. With $\omega_s$ defined as
in (\ref{SSYMPeq}), by inspection we see that
 \begin{eqnarray}
 X_{\Xmu} &=& \Dp{}{\Pmu} \End
 X_{\Pmu} &=& -\Dp{}{\Xmu}
  - \Gam{\mu}{\nu}{\rho} \pi_{\rho} \Dp{}{\pi_{\nu}}
  + \Gam{\mu}{\nu}{\rho} \eta^{\nu} \Dp{}{\eta^{\rho}}
  - \frac12 R_{\mu\nu\rho}{}^{\Lam} \pi_{\Lam}\eta^{\rho} \Dp{}{p_{\nu}}\End
 X_{\Etamu} &=& \Dp{}{\Pimu} \End
 X_{\Pimu} &=& \Dp{}{\Etamu}.
 \end{eqnarray}
Poisson brackets are then defined by the rule
\begin{equation}\label{PBRULEeq}
  \Pb{f}{g} = \frac12 \left(  X_f \Vip d g - X_g \Vip d f \right)
\end{equation}
which leads to (\ref{SPBeq}).

\section{Sign conventions for integrals}\label{SCap}

For the supermanifold $SM$ Berezin integration corresponds to
integration of top forms on $M$. If a function $f$ on $SM$ takes
the form $f(x,\eta)= f_{\mu_1 \dots \mu_p}(x) \eta^{\mu_1} \dots
\eta^{\mu_p}$ then the conventional integral
 \[
  \int_{M}  f_{\mu_1 \dots \mu_p} (x)dx^{\mu_1} \dots dx^{\mu_p}
 \]
is equal to the Berezin integral
 \[
  \int_{SM} d^{n}x d^n \eta f_{\mu_1 \dots \mu_p} (x)\eta^{\mu_1} \dots
 \eta^{\mu_p}.
 \]

If $K$ is a linear operator on functions on the supermanifold
$SM$, then the integral kernel of $K$ (if it exists) is defined by
\begin{equation}
  K f(Y) = \int_{SM} d^{n}x d^n \eta f(X) K(X, Y).
\end{equation}

\section{Morse Theory terminology}
\label{MTap} Collected together here are some basic definitions
and notation for Morse theory; more details, and many more
results, may be found in the classic book of Milnor \cite{Milnor}.

We start with a function $h:M \to \Real$.  The {\em critical
points} of $h$ are the points where all the partial derivatives
are zero. In this paper we will assume that $h$ is a {\em Morse
function}, that is, its critical points are all isolated; a
critical point $a$ of $h$ is said to be of {\em index} $p$ if the
Hessian matrix
 $\left(\Dctwoh{a}{\Xmu}{x^{\nu}}\right)$ has exactly
 $p$ negative eigenvalues.
The set of critical points of $h$ will be denoted $\Ch$. (Although
we give here a coordinate-based definition of critical point and
index, the definitions are of course intrinsic, independent of any
choice of local coordinates.)

Each critical point $a$ of $h$ has a neighbourhood  $N_a$ on which
special coordinate systems, known as {\em Morse coordinates}, can
be chosen in which the Morse function $h$ takes the standard form
 \begin{equation}\label{MFeq}
  h(x) = h(a) + \frac12 \sum_{\mu=1}^{n} \Sigm (x^{\mu}-a^{\mu})^2
 \end{equation}
 with $\Sigm =+1$ for $\mu=1, \dots, n-p$ and
 $\Sigm =-1$ for $\mu=n-p+1,\dots,n$.
(Here, abusing notation for simplicity, points and their
coordinates are identified.) On the corresponding neighbourhood of
the supermanifold $SM$ we use odd coordinates $\kappa_{\mu}$
corresponding to $dx^{\mu}$.

A metric $g$ on $M$ satisfies the Smayle transversality condition
for $h$ if the solution curves $\Gamma_{ab}$ to the 'steepest
descent'  differential equation
 \begin{equation}\label{STeq}
 \frac{dx^{\mu}(t)}{dt} = -g^{\mu\nu}\Dp{h}{x^{\nu}}
 \end{equation}
 which start from a critical point $b$ and end at a critical point $a$
(with $h(a)$ necessarily less than $h(b)$) are discrete (and
finite in number).
%
%


\begin{thebibliography}{10}
\bibitem{Witten82}
E.~Witten.
\newblock Supersymmetry and {M}orse theory.
\newblock {\em Journal of Differential Geometry}, 17:661--692, 1982.

\bibitem{BeaSin}
Beaulieu and I.~Singer.
\newblock The topological sigma model.
\newblock {\em Commun. Math. Phys.}, 125:227--237, 1989.

\bibitem{Nicola80}
H.~Nicolai.
\newblock On a new charaterization of scalar supersymmetric theories.
\newblock {\em Physics Letters}, B89:341--346, 1980.

\bibitem{BirRakTho88}
D.~Birmingham, M.~Rakowski, and G.~Thompson.
\newblock Topological field theories, {N}icolai maps and {BRST} quantization.
\newblock {\em Physics Letters}, B214:381--386, 1988.

\bibitem{BlaKesNie90}
M.~Blau, E.~Keski-Vakkuri, and A.J. Niemi.
\newblock Path integrals and geometry of trajectories.
\newblock {\em Physics Letters}, B246:92--98, 1990.

\bibitem{HenTei}
M.~Henneaux and C.~Teitelboim.
\newblock {\em Quantization of Gauge Systems}.
\newblock Princeton University Press, 1992.
\newblock in bfv.bib.

\bibitem{KosSte}
B.~Kostant and S.~Sternberg.
\newblock Symplectic reduction, {BRS} cohomology, and infinite-dimensional
  {C}lifford algebras.
\newblock {\em Annals of Physics}, 176:49--113, 1987.

\bibitem{Hennea}
M.~Henneaux.
\newblock Hamiltonian form of the path integral for theories with a gauge
  freedom.
\newblock {\em Phys. Rep.}, 126:1, 1985.

\bibitem{GFBFVQ}
A.~Rogers.
\newblock Gauge fixing and {BFV} quantization.
\newblock {\em Classical and Quantum Gravity}, 17:389--397, 2000.

\bibitem{SalHol82}
P.~Salmonson and J.W. van Holten.
\newblock Fermionic coordinates and supersymmetry in quantum mechanics.
\newblock {\em Nuclear Physics}, B196:509--531, 1982.

\bibitem{LanLif58}
L.D. Landau and E.M. Lifschitz.
\newblock {\em Quantum Mechanics}.
\newblock Pergamon Press, 1958.

\bibitem{Colema85}
S.~Coleman.
\newblock {\em Aspects of Symmetry, Selected Erice Lectures}, chapter 7: The
  Uses of Instantons.
\newblock Cambridge University Press, 1985.

\bibitem{JMPFI}
Alice Rogers.
\newblock Path integration, anticommuting variables and supersymmetry.
\newblock {\em Journal of Mathematical Physics}, 36:2531--2545, 1995.

\bibitem{Elwort}
K.D. Elworthy.
\newblock {\em Stochastic Differential Equations on Manifolds}.
\newblock London Mathematical Society Lecture Notes in Mathematics. Cambridge
  University Press, 1982.

\bibitem{IkeWat}
N.~Ikeda and S.~Watanabe.
\newblock {\em Stochastic differential equations and diffusion processes}.
\newblock North-Holland, 1981.

\bibitem{Bismut86}
J.-M. Bismut.
\newblock The {W}itten complex and the degnerate morse inequalities.
\newblock {\em Journal of Differential Geometry}, 23:207--240, 1986.

\bibitem{Simetal}
H.L. Cycon, R.G. Froese, W.~Kirsch, and B.~Simon.
\newblock {\em Schrodinger operators}.
\newblock Springer, 1987.

\bibitem{Simon}
B.~Simon.
\newblock {\em Functional Integration and Quantum mechanics}.
\newblock Academic Press, 1979.

\bibitem{GBM}
A.~Rogers.
\newblock Fermionic path integration and {G}rassmann {B}rownian motion.
\newblock {\em Communications in Mathematical Physics}, 113:353--368, 1987.

\bibitem{Hrabak}
S.P. Hrabak.
\newblock On the multisymplectic origin of the nonabelian deformation algebra
  of pseudoholomorphic embeddings into strictly almost kahler ambient
  manifolds, and the corresponding {BRST} algebra.
\newblock Preprint math-ph/9904026, 1999.

\bibitem{Witten88}
E.~Witten.
\newblock Topological sigma models.
\newblock {\em Commun.Math.Phys.}, 118:411, 1988.

\bibitem{Kanatc}
I.~V. Kanatchikov.
\newblock On quantization of field theories in polymomentum variables.
\newblock {\em To be published in the proceedings of International Conference
  on Particles, Fields and Gravitation (Devoted to the Memory of Professor
  Ryszard Raczka), Lodz, Poland, 15-18 Apr 1998.}, hep-th/9811016 AIP
  Proceedings 1998.

\bibitem{Milnor}
Milnor.
\newblock {\em Morse Theory}.
\newblock Princeton University Press, 1963.

\end{thebibliography}
\end{document}